\documentclass[twocolumn,prb,aps,floats,showpacs]{revtex4-1}
\usepackage{amsmath}

\usepackage{epsfig}
\usepackage{graphicx}
\usepackage{dcolumn}
\usepackage{bm}
\usepackage{amssymb}
\usepackage[utf8]{inputenc}

\def\gapp{\lower.35em\hbox{$\stackrel{\textstyle>}{\sim}$}}
\def\lapp{\lower.35em\hbox{$\stackrel{\textstyle<}{\sim}$}}

\begin{document}
\bibliographystyle{apsrev4-1}
%

\title{Spectral and optical properties of doped graphene with charged impurities
in the self-consistent Born approximation}

\author{F. de Juan$^1$, E.H. Hwang$^2$, M.A.H. Vozmediano$^1$}

\affiliation{$^1$ Instituto de Ciencia de Materiales de Madrid,\\
CSIC, Cantoblanco, E-28049 Madrid, Spain.}

\affiliation{$^2$ Condensed Matter Theory Center, Department of Physics, \\
University of Maryland, College Park, Maryland 20742-4111}




\date{\today}
\begin{abstract}
Spectral and transport properties of doped (or gated) graphene with
long range charged impurities are discussed within the self-consistent Born
approximation. It is shown how,
for impurity concentrations $n_{imp}\gtrsim n$ a finite DOS appears at the Dirac
point, the one-particle lifetime no longer scales linearly with the Fermi
momentum, and the lineshapes in the spectral function become non-lorentzian.
These behaviors are different from the results
calculated within the Born approximation. We also calculate the optical
conductivity
from the Kubo formula by using the self-consistently calculated
spectral function in the presence of charged impurities.

\end{abstract}
%
%
%
%

\maketitle

\section{Introduction}

Graphene, the monolayer allotrope of carbon, has attracted widespread
attention since its isolation \cite{NGM04,NGM05}, and remains to be
the focus of intensive research. Among the main reasons for this
interest are its remarkable electronic properties, described in terms
of massless Dirac fermions. They make graphene an appealing system to
study new unconventional physics \cite{CGP09}, but moreover, easy
control of the electron density through gating \cite{NGM04} and very
high room temperature mobilities \cite{G09} make it also a promising
material for applications.  
Since the first experiments, it was acknowledged that understanding the role of
disorder was essential
to describe the electronic properties of graphene correctly
\cite{P09,AAB10,DAH10,ML10}. 
One of the first issues raised and still under debate is the origin of the main
scattering source in
graphene transport.
The linear-in-density dependence of the DC conductivity \cite{NGM05} 
was first attributed to charged impurities trapped in the substrate \cite{A06a,HAD07,CF06,NM06},
which were considered the main scattering mechanism. But it was
later shown that resonant scatterers \cite{SPG07}
and ripples \cite{KG08} could also account for this linear behavior. 
Although Coulomb impurities are inevitably present in graphene samples
and influence the transport properties, they need not always be the
dominant scattering mechanism limiting the mobility especially at high
densities \cite{NPN10}.

The minimal conductivity measured at the Dirac point is 
another experimental observation which can clarify the main scattering
mechanism in graphene \cite{CGP09}. While the universal value of
$4e^2/\pi h$ was predicted for uncorrelated short range disorder as well as for
ripples (described in terms of random gauge fields), the experimental value was
consistently found larger and sample dependent \cite{NGM05,TZB07}. 
These experimental observations can be explained by transport theory in the
presence of Coulomb disorder and electron-hole puddle
formation \cite{NM07,DAH10,P10a}. 
DC transport measurements have therefore provided compelling evidence of the
relevance of charged impurities in the physics of graphene in substrates. 

The comparison of the single particle relaxation time $\tau_{e}$ defining the quantum
level broadening and the transport scattering time $\tau_{tr}$ defining the
Drude conductivity in 2D graphene layers 
can also be a relevant probe of the nature of disorder
scattering of graphene carriers \cite{HD08a}.
Hong {\it et al}. \cite{HZZ09} report, by comparing
these two independently measured scattering times, that Coulomb impurities play
a dominant role in graphene samples.
But another recent experiment of the same type suggests
that the main scattering mechanism in graphene
is due to strong (resonant) scatterers of a range shorter
than the Fermi wavelength \cite{MOW10}.

The study of the local density of states in samples on a substrate \cite{ZBG09}
has
revealed inhomogeneites which have been also attributed to charged impurities
\cite{DAH10} although ripples may also contribute to this phenomenon
\cite{JCV07}. Long range charged disorder may
also contribute to the
broadening of the spectral linewidth observed in angle-resolved
photoemission experiments \cite{ZGF07,BOS07,KWM08}.
 
The long range charged disorder is also
important to understand the recently measured optical conductivity
\cite{LHJ08,HCG10}. These experiments have revealed deviations from the ideal
picture of Dirac fermions which have been attributed, at least in part, to the
presence of Coulomb
impurities \cite{SPC08}. 

Thus it is expected that Coulomb disorder inevitably existing in the environment
of graphene will be important in many physical properties. In this paper we
investigate
spectral and transport properties of doped (or gated) graphene with
long range charged impurities within the self-consistent Born approximation. 

In general, the Born approximation has been widely used to treat the
Coulomb disorder \cite{R98}. However, the first order Born approximation is 
restricted by  $n_{imp} \ll n$,
which is not necessarily satisfied in most experiments. More
elaborate numerical approaches \cite{NM07,QAA08,DAH10,P10a} are not
hindered by this restriction, but the simplicity of the averaged
theory makes it appealing to extend the known results beyond first
order.  
The self-consistent Born approximation (SCBA) is a non-perturbative
approximation \cite{R98}, and it is the simplest natural way to extend
the Born approximation. In the case of graphene, this approximation has been used mainly for
momentum independent $\delta$-ranged impurities. The more complicated case of
Coulomb impurities in SCBA has been addressed in the evaluation of the DC conductivity
only \cite{YT08,YRT08}. In this article, we present the numerical solution of
the SCBA equations for doped (or gated) graphene, and compute both
spectral and
optical properties by using the calculated self-energy and compare our
results with previous works and with experiments. We observe important changes
for impurity 
concentrations $n_{imp}\gtrsim n$, as compared to the first order case: a finite
DOS at the Dirac
point, non-linear one-particle lifetimes, non-lorentzian spectral functions, and
a modified optical conductivity.

The organization of the paper is as follows: we start by reviewing the
single Coulomb impurity problem in graphene in section
\ref{impurity}. In section \ref{model} we discuss the treatment of
random Coulomb impurities in an averaged theory, and formulate the
SCBA equations. Section \ref{results} describes the results of our
work: spectral properties and the optical conductivity are
discussed. Finally, we present our conclusions in section
\ref{discussion}.  

\section{The Coulomb impurity problem in graphene}\label{impurity}

The problem with a single Coulomb
impurity has been widely
studied in both low density limit 
\cite{DM84,FNS07,TMK08,K06c,SKL07,BSS07,PNC07} and high density
limit \cite{K06c,GMG09}. This problem is the starting point in the
description of Coulomb disordered graphene, so we
highlight its most relevant aspects.  
The importance of the single impurity problem stems from the fact that
an electron in graphene is affected not only by the bare Coulomb
potential of the impurity, but also by the other electrons that
redistribute around it. The effective impurity potential is thus
screened by the carriers, and the inclusion of the screening is very important
to
account correctly for the effects of a random collection of
impurities. Since the exact consideration of the screened Coulomb
potential is a difficult problem, the screening is simplified 
by assuming a linear 
response approach. In this linear screening limit, the screened
Coulomb potential is given by
\begin{equation}
V(q) = \frac{V_{0}(q)}{\epsilon(q,\omega \rightarrow 0)},
\end{equation}
where $\epsilon(q,\omega)$ is the dynamical dielectric function and
can be obtained from the density-density correlation function as
\begin{equation}
\frac{1}{\epsilon(q,\omega)}=1+V_{0}(q)\Pi(q,\omega),
\end{equation}
where $\Pi(q,\omega)$ is the polarizability of the system. 
In the weak interaction limit the $\Pi(q,\omega)$ is calculated within
random phase approximation (RPA) by summing all bare bubbles. 

In ordinary 2DEG the weak interaction limit is
defined by the interaction parameter $r_{s}<<1$, where $r_{s}$ is given by 
\begin{equation}
r_{s}=\frac{m e^2}{(\pi n )^{1/2}},
\end{equation}
where $n$ is the 2D electron density and $m$ is the electron effective
mass \cite{AFS82} .
If the density is lowered, non-linear response 
comes into play \cite{ZNE03}, and the parameter $r_{s}$
determines the range of densities where the linear RPA model is applicable. 

In the case of graphene, one could expect that the RPA is a good
approximation at high densities but its validity near Dirac point is
questionable. 
In contrast to the regular
2DEG, the parameter $r_{s}$ of graphene is density independent, and it
is known as 
the coupling constant $\alpha=e^{2}/(\kappa v_{F})$. This generates a
difficulty in evaluating the range of validity of the RPA for graphene. It is
understood that in the zero doping case it is definitely not valid \cite{GFM08},
while it has been stated that in the doped case it is applicable when
$r_{s} \ll 1$ \cite{DHH07}, the same criterion for the use of normal
perturbation theory. Since $r_{s}$ is density independent, it is not
clear how the two regimes interpolate, and to the best of our
knowledge no quantitative criterion on $n=k_{F}^{2}/\pi$ has been
established to separate them. 

In spite of that, due to its simplicity and its qualitative success in
comparison with
experiments, the RPA for graphene has been thoroughly studied in the
literature \cite{CGP09}. In static limit the
polarizability is given by \cite{HD07}
\begin{equation}\label{epsilon}
\epsilon(q)=1+\frac{q_{TF}}{q}\Pi(\frac{q}{k_{F}}),
\end{equation}
where $q_{TF} =  4 \alpha_{v} k_{F}$ and
\begin{equation}\label{pi}
\Pi(x)=\left\lbrace 
\begin{array}{cc}
1 & x<2 \\
1+\frac{\pi x}{8} -\frac{\sqrt{x^2-4}}{2x}-\frac{x \arcsin(2/x)}{4} & x>2
\end{array}
\right. .
\end{equation}
For $k \le 2k_{F}$, this is just the Thomas-Fermi result \cite{K06c},
and identical to the regular 2DEG \cite{AFS82}. Thus the difference in
screening between 2DEG and graphene arises at high wave vectors, $k>2k_{F}$. 
Even though the high density
screening is similar to the 2DEG, it is worth noting that 
as $n \rightarrow 0$ RPA screening in graphene only changes the
constant dielectric
constant, but not the wave vector dependence. This
is due to the vanishing density of states at the Fermi level as $n
\rightarrow 0$. 
The RPA result is questionable at $n=0$, where the
problem becomes strongly non-linear
\cite{DM84,FNS07,TMK08,K06c,SKL07,BSS07,PNC07,SK10}, but its results
compare favorably with tight binding exact results \cite{BF09}. In
this work, both the linear response RPA polarizability and the
Thomas-Fermi result will be used as 
a model of screening, bearing in mind the previous discussion
regarding their limits of applicability.  

Finally, it is also worth noting that when we consider
many impurities, we face a more complex problem in terms of
screening (even before disorder averaging), because the 
screening of one impurity may depend on the charge accumulated on the
rest of them. 
It is reasonable to assume independent screening of impurities 
when the impurity density is small,
but this picture may fail when the screening length becomes
bigger than the average distance between impurities $n_{imp}^{-1/2}$. The many
impurity problem introduces another type of non-linearity which may become
relevant at low densities, where the screening is weak
\cite{E88a}. Since the Thomas-Fermi screening length is given by  
\begin{equation}
l_{TF} = q_{TF}^{-1} = \frac{1}{4\alpha (\pi n)^{1/2}}, 
\end{equation}
the condition for independent screening can be written as
\begin{equation}\label{indsc}
n_{imp}\lesssim (4\alpha)^2 \pi n. 
\end{equation}

In the case of graphene on SiO$_2$, we can take $\alpha \sim 0.75$ and this means $n_{imp}/n
\lesssim 27$ which is well satisfied for the ranges of $n$ and $n_{imp}$ we will consider. 

\section{Averaged theory for random charged impurities}\label{model}

In the standard theory of disorder \cite{R98}
is is assumed that
the properties of the system with a particular
realization of the disorder landscape are the same as those averaged
over all impurity positions. 
In the case of graphene, the low energy Hamiltonian
for a particular distribution of $N_{imp}$ impurities is given by
\begin{equation}
H = \int d^{2}r \left[ v_{F} \psi^{\dagger}\boldsymbol\sigma
\boldsymbol\partial\psi - \mu \psi^{\dagger}\psi +\psi^{\dagger}V(r)\psi
\right],
\end{equation}
where $v_{F} = 10^6$ m/s is the Fermi velocity, and $\mu$ is the chemical
potential (note that energies are measured with respect to the Dirac
point). This Hamiltonian is a good approximation to the band structure
for energies up to $E \lesssim$ 1 eV \cite{CGP09}, so we will consider the properties of this
model in this range of energy. The largest electron density we will consider will be $n = 6
\cdot  10^{12}$ cm$^{-2}$, which corresponds to a chemical potential of $\mu =$ 0.3 eV, well below
the limit of applicability. The disorder potential $V(r)$ is given by 
\begin{equation}
V(r) = \sum_{i}^{N_{imp}} \int dq e^{i(r-R_{i})q} V(q),
\end{equation}
where $R_{i}$ are the positions of the $N_{imp}$ impurities, and
$V(q)$ is the Fourier transform of the potential.  The disordered
system is described by its Green functions averaged over impurity
positions, which we assume uncorrelated for simplicity. 
In the case of graphene, this approach has been widely used to model
disorder \cite{HHD08,PGC06,YPX10,YRT08}. When the potential is
weak, multiple scattering events can be neglected, and the resulting
perturbative series is known as the Born approximation, of which only
the first terms are necessary. The first order term in 
this series for a general potential reads 
\begin{equation}\label{fob}
\Sigma_{s}(k,\omega) = n_{imp} \int
\frac{dk^2}{(2\pi)^2}V^2(k-k')\sum_{s'}
G_{0,s'}(k',\omega)F_{ss'}(k,k'), 
\end{equation}
where $s=+,-$ for the upper and lower bands, $n_{imp}$ is the density of
impurities
\begin{equation}
n_{imp}=N_{imp}/L^2,
\end{equation}
with $L$ the system size. $G_{0,s'}$ are the bare Green functions for each
band
\begin{equation}
G_{0,\pm}(k,\omega) = \frac{1}{\omega \pm v_{F} |k| +i\epsilon}
\end{equation}
and $F_{ss'}$ are the overlap factors
\begin{equation}
F_{ss'} = \frac{1}{2}(1+ss'\cos \theta_{k,k'}).
\end{equation}

In the self-consistent Born approximation the imaginary part of the
self-energy in the bare Green function is
replaced by the self energy of the full Green function, i.e., 
$i\epsilon \rightarrow \Sigma$, and we have the self-consistent
equation as
\begin{equation}
G^{-1}_{s} = G_{0,s}^{-1} -\Sigma_{s}
\end{equation}
where the self energy is given by
\begin{equation}\label{scbagen}
\Sigma_{s}(k,\omega) = n_{imp} \int
\frac{dk^2}{(2\pi)^2}V^2(k-k')\sum_{s'}
G_{s'}(k',\omega)F_{ss'}(k,k'). 
\end{equation}
The SCBA for the short range case gives well known results. Note that, since the
potential is independent of $k$, the self-energy in this approximation is also
independent of $k$. At $\omega=0$ it can be solved analytically, giving the well
known finite purely imaginary self-energy
\begin{equation}\label{sigmashort}
\Sigma(\omega=0)=-\frac{i\Lambda}{\sqrt{\exp(\frac{4\pi
v_{F}^2}{n_{imp}V_{0}^{2}})-1}},
\end{equation}
and the DOS
\begin{equation}\label{firstrho}
\rho_{0}=\frac{1}{\pi n_{imp}V_{0}^2}\frac{\Lambda v_{F}}{\sqrt{\exp(\frac{4\pi
v_{F}^{2}}{n_{imp}V_{0}^{2}})-1}}.
\end{equation}
where $\Lambda$ is a high energy cutoff. For undoped graphene, however, it has been long known that
the SCBA is in
principle not a good approximation due to the absence of a $E_{F}\tau$ type of
parameter that
would allow to neglect the crossing diagrams \cite{NTW95,AE06} (but
see ref. \onlinecite{F86a} for a special case where the non-crossing
approximation is controlled by 1/N, N being the number of valleys). 
Renormalization group calculations as well as exact results
\cite{ASZ02} indeed differ
qualitatively from the SCBA.  

Nevertheless, it still represents a good approximation in the doped
case. While the short range case is analytically tractable, the SCBA
has not been applied to Coulomb impurities because its implementation
is not as simple \cite{K07}. The potential due to
screened Coulomb impurities is given by 
\begin{equation}\label{coulomb}
V_{C}(q)=\frac{2\pi e^2}{\kappa \epsilon(q)}\frac{1}{q},
\end{equation}
as stated in the previous section, and the dielectric function is
given by Eqs.~(\ref{epsilon}) and (\ref{pi}) (for comparison, we will also
use the Thomas-Fermi dielectric function, the low $q$ limit of Eq.~(\ref{pi})). 

In the first order Born approximation, the imaginary part of the
self-energy has been obtained analytically.
The computation of Eq. (\ref{fob}) with the
Coulomb potential Eq. (\ref{coulomb}) gives, at k=0 \cite{SPC08}:  
\begin{equation}
Im[\Sigma^{(1)}(0,\omega)] = n_{imp} \frac{e^4}{\kappa^2} \int k' dk' d\theta
\frac{\pi \delta(\omega-v_{F}k')}{(|k'|+q_{TF})^2}\frac{1+\cos\theta}{2}, 
\end{equation}
which gives
\begin{equation}\label{firstk0}
Im[\Sigma^{(1)}(0,\omega)] = n_{imp}\pi^2 \alpha^2  \frac{\omega}{(\omega/v_{F}+q_{TF})^2}.
\end{equation}
Since this was computed in the Thomas-Fermi approximation, it is only
valid for $\omega<2\omega_F$. At $k=k_F,\omega = \omega_F$ we have
\cite{HD08a} 
\begin{equation}\label{firstborn}
\Sigma = \frac{\alpha^{2} n_{imp} v_{F} \pi I(2\alpha)}{k_F},
\end{equation}
where $I(2\alpha)$ is a dimensionless function which can also be found
in ref. \onlinecite{HD08a} (for our purposes we will take $\alpha = 0.75$
and we have $I(2*0.75) = 
0.224$). This self-energy allowed to compute the density of states in this
approximation, which was shown to vanish as $|E| \ln |E|$ at the Dirac point
\cite{HD08a}. 
The SCBA equations for Dirac fermions in the presence of Coulomb
impurities are obtained by substitution of the Coulomb potential
Eq. (\ref{coulomb}) in Eq. (\ref{scbagen}), obtaining explicitly 
\begin{widetext}
\begin{align}\label{scba}
\Sigma_{+}(k,\omega)= n_{imp} \int \frac{d^{2}k'}{(2\pi)^2} \left[
V_{C}(k-k')\right] ^2 \frac{1}{2}
\left[\frac{1+\cos\theta}{\omega-v_{F}k'-\Sigma_{+}(k',\omega)}
+\frac{1-\cos\theta}{\omega+v_{F}k'-\Sigma_{-}(k',\omega)}\right], \\
\Sigma_{-}(k,\omega)= n_{imp}\int \frac{d^{2}k'}{(2\pi)^2}  \left[
V_{C}(k-k')\right]
^2 \frac{1}{2}\left[\frac{1-\cos\theta}{\omega-v_{F}k'-\Sigma_{+}(k',\omega)}
+\frac{1+\cos\theta}{\omega+v_{F}k'-\Sigma_{-}(k',\omega)}\right].
\end{align}
If all quantities with dimensions are scaled with the Fermi momentum
or energy, $x=k/k_F$, $y=\omega/\omega_{F}$, and $\widetilde{\Sigma} =
\Sigma/\omega_{F}$ this formula can be rewritten as  
\begin{equation}\label{dimensionless}
\widetilde{\Sigma}_{+}(x,y)= \frac{\alpha^2}{\pi}\frac{n_{imp}}{n} \int x'dx'
d\theta \left[
\frac{1}{|x-x'|+\frac{q_{TF}}{k_{F}}}\right] ^2 \frac{1}{2}
\left[\frac{1+\cos\theta}{y-v_{F}x'-\widetilde{\Sigma}_{+}(x',y)}
+\frac{1-\cos\theta}{y+v_{F}x'-\widetilde{\Sigma}_{-}(x',y)}\right],
\end{equation}
\end{widetext}
with
\begin{equation}
k_{F}=(n \pi)^{1/2}.
\end{equation}
The iteration of formula Eq. (\ref{dimensionless}) generates the
series of diagrams for the SCBA, and therefore the parameter 
\begin{equation}
\gamma = \frac{\alpha^2}{\pi}\frac{n_{imp}}{n},
\label{gamma}
\end{equation}
controls the relevance of higher order terms. Self consistent effects
become important when $\gamma \simeq 1$. Note that the independent
screening condition Eq. (\ref{indsc}) in terms of this parameter reads
$\gamma < (2\alpha)^4 \simeq 5$. In real experiments, doping through a
gate allows to reach values of $n$ as high as $10^{13}$ cm$^{-2}$
(see ref. \onlinecite{NGM04}), and the concentration of impurities in samples is
estimated to reach up to $n_{imp}=5\cdot 10^{12}$ cm$^{-2}$ in the
most disordered ones \cite{CJA08,TZB07}.  

The solution of Eq. (\ref{dimensionless}) can be obtained numerically
by discretizing $k'$ and iterating it until an error bound is
reached. Since the momentum lattice has to be kept fixed so the
self-energy from one iteration can 
be fed to the next one, a simple Simpson rule for the integration
proved to be the most efficient way (interpolating in k-space at each
step allowed for more efficient integration 
algorithms, but the overall performance of this strategy turned out to be worse
than the simple Simpson rule). For the integration in k-space to be reliable,
the step in the
discretization has to be much smaller than the imaginary part of the
self-energy, because this determines the size over which the function to
integrate is significantly different from zero. This sets a practical
limitation to the lowest value of $n_{imp}/n$ to which we have access,
which is of the order of 0.1.  

Once the self-energy has been obtained, the spectral properties of the
system are easily computable. The density of 
states is obtained from
\begin{equation}\label{dosst}
\rho(\omega) = \frac{1}{\pi} \lim_{x\rightarrow x'} Im[G(x,x',\omega)].
\end{equation}
The inverse quantum lifetime is defined as $\tau_{e}^{-1}(k,\omega) =
Im[\Sigma(k,\omega)]$, and it is in general a function of both
$\omega$ and $k$. Its most 
interesting value is the on-shell lifetime at the Fermi energy
$\tau_{e}^{-1}(k_{F},E_{F})$. Finally, the spectral function is computed
as the imaginary part of the Green function 
\begin{equation}\label{spec1}
A(k,\omega) = \frac{1}{\pi}\left[ A_{+}(k,\omega)+A_{-}(k,\omega)\right],
\end{equation}
with 
\begin{equation}\label{spec2}
A_{+}(k,\omega)= \frac{Im[\Sigma_{+}(k,\omega)]}{(\omega -
  \omega_{+}(k) -
  Re[\Sigma_{+}(k,\omega)])^2+Im[\Sigma_{+}(k,\omega)]^2}, 
\end{equation}
and similarly for $A_{-}(k,\omega)$, where $\omega_{\pm}(k) = \pm
v_{F}|k|$.

\section{Results}\label{results}

\subsection{Spectral properties}\label{spectralsec}

\begin{figure}[h]
\begin{center}
\includegraphics[width=8cm]{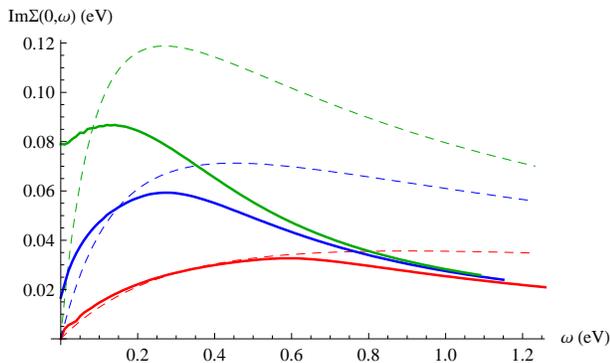}
\caption{(Color online) Imaginary part of the zero-momentum
  self-energy, $\Sigma(0,\omega)$, for $n_{imp}=6\cdot10^{12}$
  cm$^{-2}$ and different densities, $n = 6\cdot10^{11}$,
  $1.7\cdot10^{12}$,  $6.7\cdot 10^{12}$ cm$^{-2}$, from top to bottom
  (solid lines). For comparison, the first order analytical result
  given by Eq.~(\ref{firstk0}) is also plotted for the same values of
  the parameters (dashed lines).}\label{k0sigmaimage} 
\end{center}
\end{figure}

We now discuss the self-energies and related spectral properties
obtained from Eq.~(\ref{scba}). We start from the simple case of
the zero momentum self-energy, $\Sigma(k=0,\omega)$, often taken as an
approximation for the self-energy around the Dirac
point. Fig.~\ref{k0sigmaimage} shows the imaginary part of the
zero-momentum self-energy for $n_{imp}=6 \cdot 10^{12}$ cm$^{-2}$ and
several densities. For comparison, the first order Born approximation
results (i.e.,
Eq.~(\ref{firstk0})) are also plotted for the same values of the
parameters.  

Several features should be noted in Fig.~\ref{k0sigmaimage}. When
$n_{imp} \ll n$, the SCBA result agree with the first order Born approximation
result
as expected. But when the impurity density is comparable with electron
density, $n \simeq n_{imp}$, the disagreement between BA and SCBA 
is measurable, especially at high energies.
When $n_{imp}$ is further increased ($n_{imp} > n$), the
deviation becomes significant, and more interestingly a finite
value of self energy is developed
at zero energy, which gives rise to a finite density
of states at the Dirac point (see below). The finite density of states
at zero energy is one of the main
characteristics of the SCBA result as compared to the first order
Born approximation, which predicts a vanishing DOS at zero energy. 

\begin{figure}[h]
\begin{center}
\includegraphics[width=8cm]{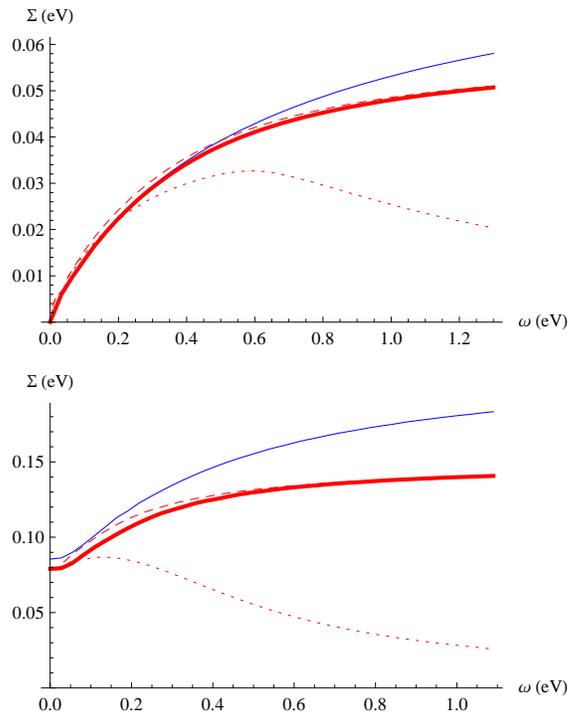}
\caption{(Color online) Different approximations for the self-energy,
  at fixed impurity concentration of $6 \cdot 10^{12}$ cm$^{-2}$, and
  for different electron densities, $6.7 \cdot 10^{12}$ cm$^{-2}$ (upper panel) and $6
  \cdot 10^{11}$ cm$^{-2}$ (lower panel). The self-energies are:
  $\Sigma(k,\epsilon(k))$ (thick line), $\Sigma(k,k)$ (dashed line),
  $\Sigma(0,\omega)$ (dotted line). $\Sigma^{TF}(k,\epsilon(k))$, as
  computed with the Thomas-Fermi approximation, is shown for
  comparison in blue (upper curve).}\label{compsigma} 
\end{center}
\end{figure}

As discussed in the introduction, 
the self-energy for the short range
potentials does not depend on the wave vector. Therefore, the
simple calculation at $k=0$ can be used for all wave vectors.
However, when the self energy is a function of the wave vector and
energy as for Coulomb disorder, the $k = 0$ self-energy may be useful to
analyze the physics near the Dirac point at high densities but, in general, 
it is not a good approximation to analyze it at finite wave vectors.
Instead of zero wave vector self-energy, the most relevant value of the
self-energy is at the on-shell self energy, $\Sigma(k,\epsilon(k))$, 
where the Green functions are peaked.
A first approximation to the on-shell self-energy is obtained simply
by taking $\Sigma(k,v_Fk)$, but the true
on-shell self-energy should be computed with the renormalized
dispersion relation, $\Sigma(k,\epsilon(k))$ (this is only relevant
when the dispersion relation is greatly changed by disorder).

The imaginary part of the three different self-energies
$\Sigma(k,\epsilon(k)),\Sigma(k,v_Fk),\Sigma(0,\omega)$ is displayed in
Fig.~\ref{compsigma} for two different impurity concentrations.  
We find that within SCBA the calculated self-energy
for all different cases shows a finite value at the Dirac point for $n_{imp} >
n$. 
Aside from this,
we observe that the $k = 0$ self-energy is indeed very
different from the on-shell self energy. One should therefore be
cautious when applying this approximation for general
computations. We also observe that the true on-shell
self-energy $\Sigma(k,\epsilon(k))$ presents almost no difference from
the approximation $\Sigma(k,v_Fk)$, which can thus be employed safely.  
Another approximation commonly used in the literature
is to employ Thomas-Fermi screening for the Coulomb impurities instead
of the full RPA dielectric 
function. We also show in Fig.~\ref{compsigma} the imaginary part of self-energy
$\Sigma(k,v_Fk)$ with the TF screening function. A notable difference is
observed
compared with the RPA calculation. In particular, note that these two curves
differ even for $\omega < 2 v_F k_F$. 

\begin{figure}[h]
\begin{center}
\includegraphics[width=9cm]{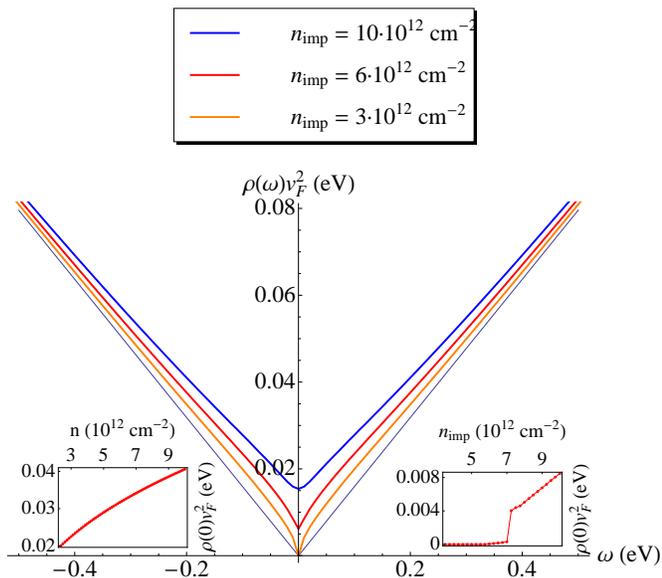}
\caption{(Color online) Densities of states as a function of energy,
  for a fixed doping $n = 1.6 \cdot 10^{12}$ cm$^{-2}$, and varying
  impurity concentration, $n_{imp} = 3 \cdot 10^{12}$, $6 \cdot
  10^{12}$, $10 \cdot 10^{12}$ cm$^{-2}$. The linear behavior for the
  clean case is shown for comparison (thin line) Left inset: Variation
  of the zero energy DOS upon simultaneous increase of $n$ and
  $n_{imp}$ (see text). Right inset: Zero energy DOS as a function of
  $n_{imp}$, for fixed doping of $n = 3 \cdot 10^{12}$
  cm$^{-2}$. }\label{dos} 
\end{center}
\end{figure}

The density of states of the system can be calculated with the self energy, 
Eq.~(\ref{dosst}). Fig.~\ref{dos} shows the densities of states as a
function of energy for a fixed electron density $n$ and several impurity
concentrations. A significant finite value is observed at the Dirac
point for high impurity concentrations, in agreement with the previous
discussion on the self-energy. To quantify better where the onset of
this finite value occurs, in the inset of Fig.~\ref{dos} we show the
DOS at the Dirac point as a function of $n_{imp}$, for $n = 3 \cdot
10^{12}$ cm$^{-2}$. A clear threshold is observed for a value of
$\gamma \simeq 0.4$. 
This finite value of the density of states is also similar to the one
obtained in the short range case. However, there is a significant
difference. In the short range case, the analytical calculation
predicts that this finite value is proportional to the cut-off, the
only scale with dimensions at $\omega=0$. The density of states,
therefore, remains constant if the rest of the scales in the problem
are changed simultaneously. To check weather this behavior is present
in the case of Coulomb impurities, we computed the zero energy density
of states for $n_{imp} = 6n$ sweeping the value of $n$. This is shown
in the left inset of fig. \ref{dos}. A clear monotonous behavior is
observed, indicating a more complicated dependence on these scales
than the constant behavior of the short range case.

\begin{figure}[h]
\begin{center}
\includegraphics[width=8.5cm]{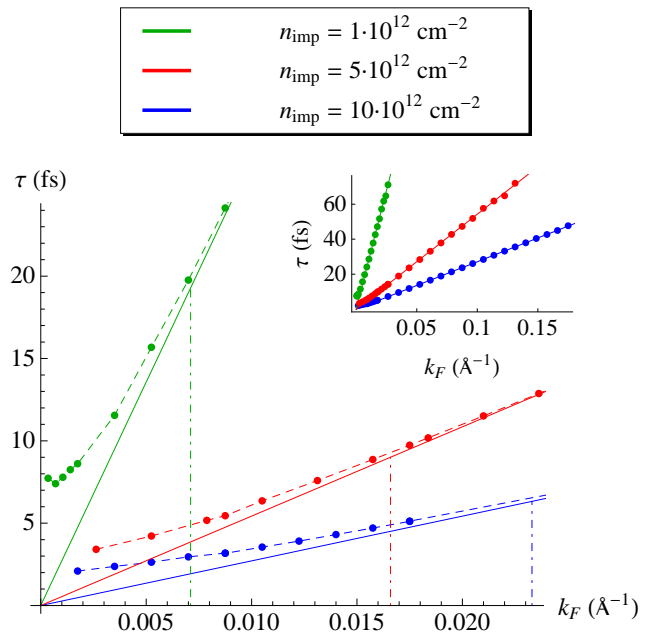}
\caption{(Color online) One particle lifetime $\tau_{e}$ as a function of $k_{F}$, for
different
impurity concentrations (the physical value of $\hbar$ is restored in
the units). The full line is the first order prediction given by 
Eq.~(\ref{firstborn}). The dashed line is a guide to the eye. A vertical
dotted-dashed line is plotted at the value $k_{F}=\alpha
\sqrt{n_{imp}}$ which corresponds to $\gamma=1$, for each
$n_{imp}$. Inset: Same plot for larger values of $k_F$, where the
linear behavior predicted by Eq.~(\ref{firstborn}) is fully
appreciated.}\label{tau} 
\end{center}
\end{figure}

Another spectral property of interest is the one particle relaxation
time $\tau_{e}$, given as the inverse imaginary part of the on-shell
self-energy at the Fermi level. Fig. 
\ref{tau} shows a plot of the dependence of $\tau_{e}$ with the Fermi level
$k_{F}$ and the first order Born approximation results, Eq.~(\ref{firstborn}). For
each concentration of impurities, the fermi momentum
$k_{F}=\alpha \sqrt{n_{imp}}$ which corresponds to $\gamma=1$ and separates the
high and low doping
region is plotted as a horizontal line. It is clearly seen that in the high
doping
region, the SCBA is again indistinguishable from the first order
result. We also see how in the low doping region, deviations from the
linear behavior are clearly identified.  

\begin{figure*}
\begin{minipage}{.49\linewidth}
\begin{center}
\includegraphics[scale=0.163]{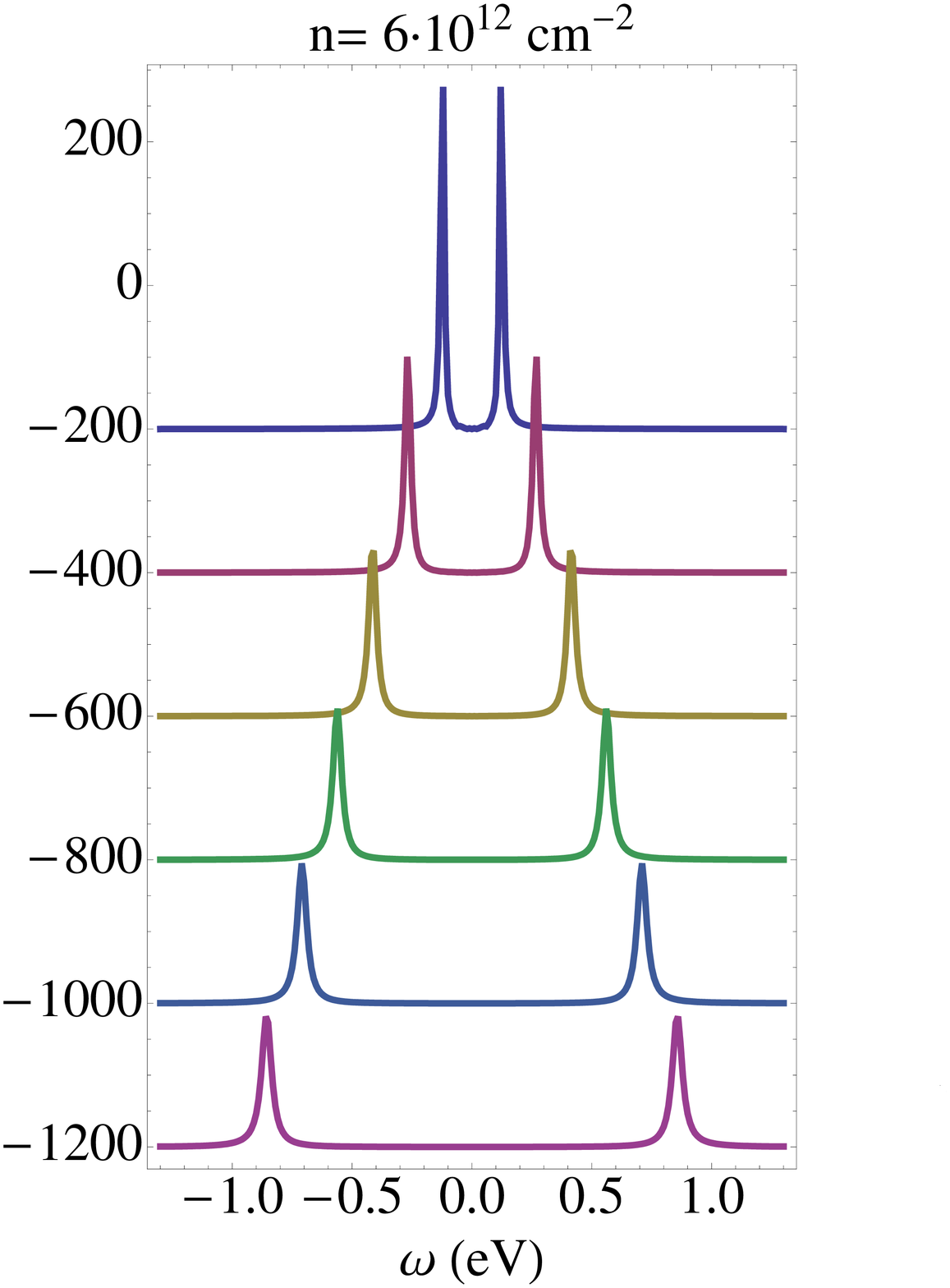}
\includegraphics[scale=0.209]{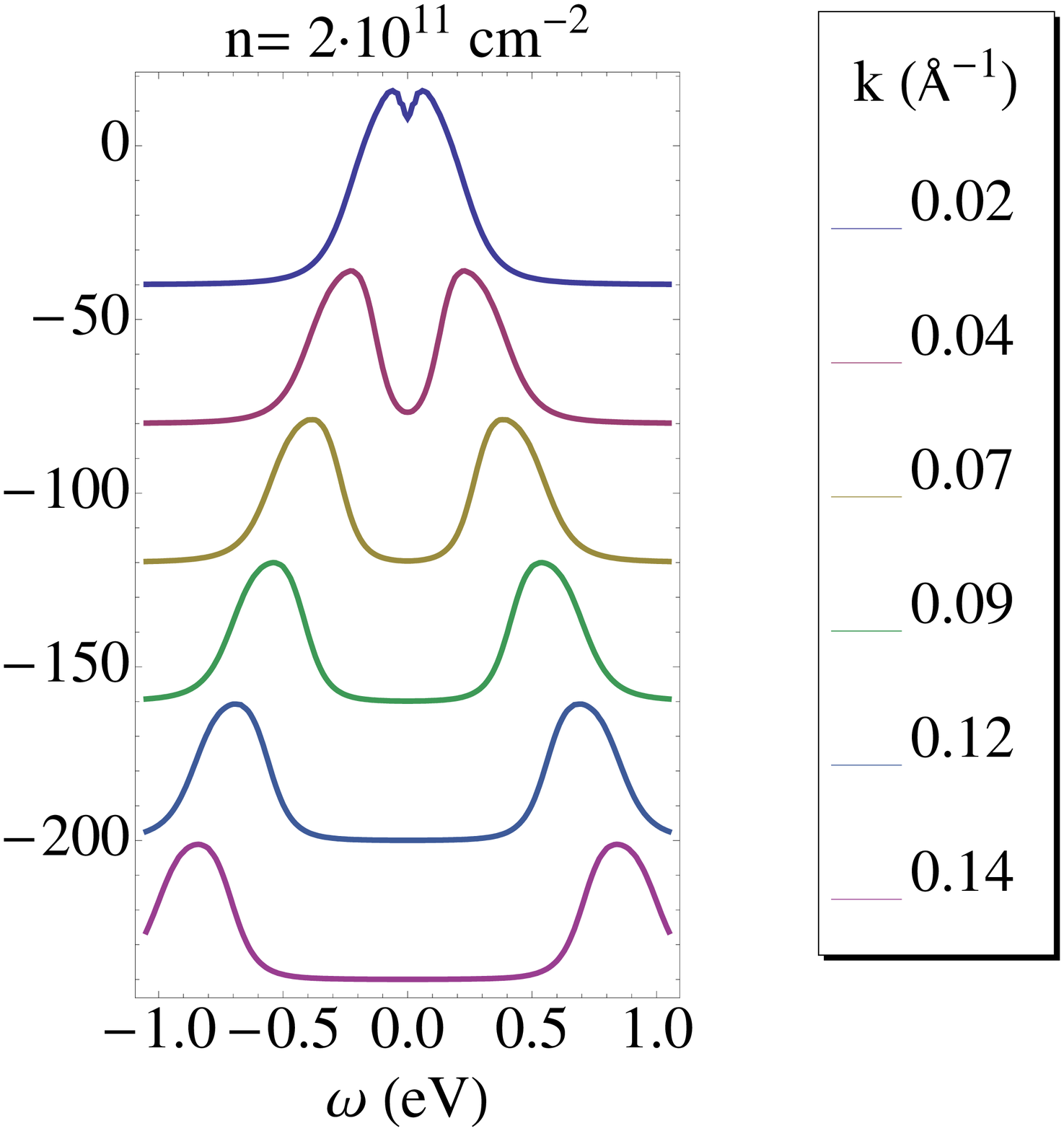}
\begin{center}
(a)
\end{center}
\end{center}
\end{minipage}
\begin{minipage}{.49\linewidth}
\begin{center}
\includegraphics[scale=0.166]{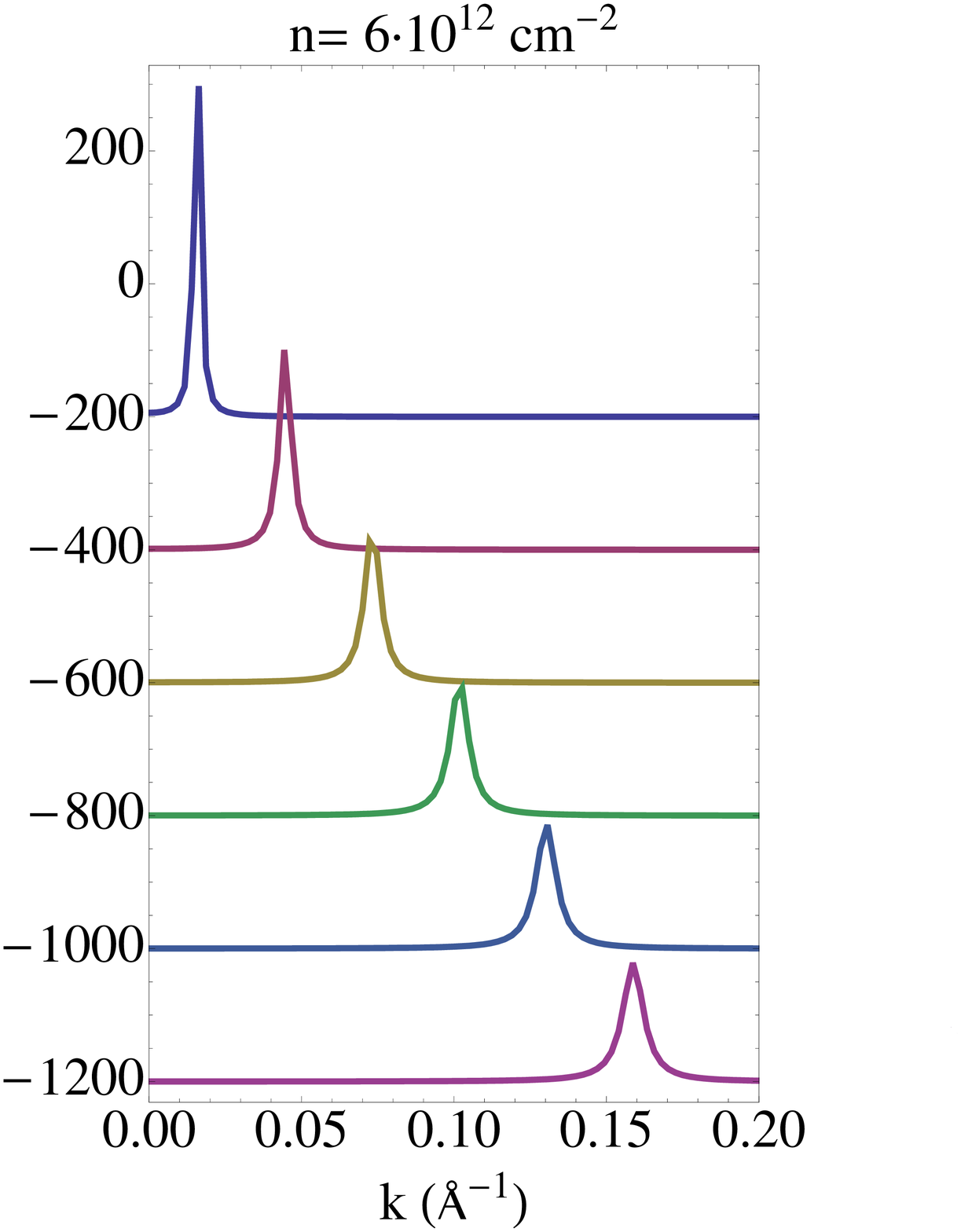}
\includegraphics[scale=0.211]{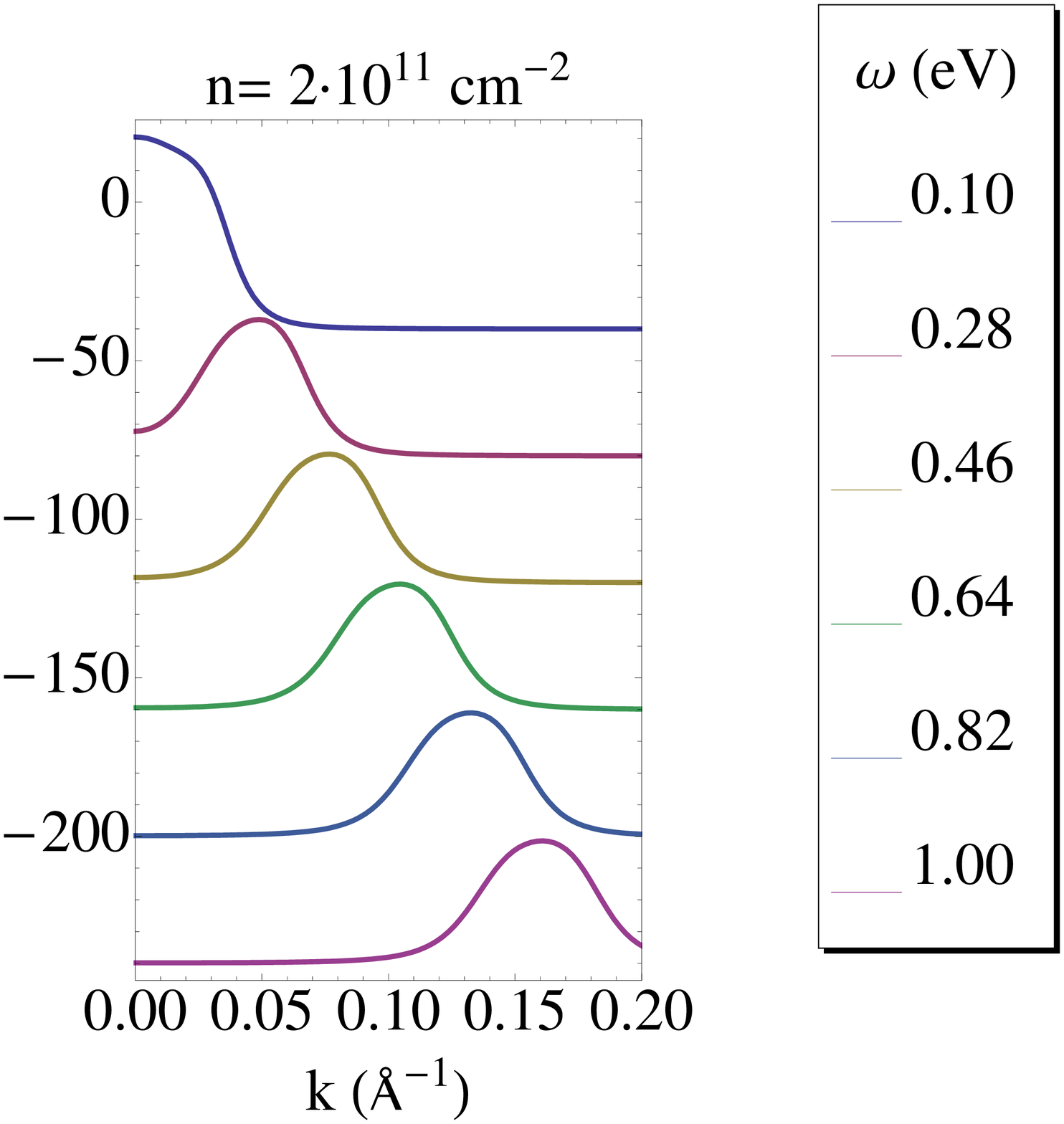}
\begin{center}
(b)
\end{center}
\end{center}
\end{minipage}\\
\caption{ (Color online) a) EDC at a fixed impurity density of $3
  \cdot 10^{12}$ cm$^{-2}$, for two values of the doping $n = 6\cdot
  10^{12}$, $2\cdot 10^{11}$ cm$^{-2}$. b) MDCs for the same parameters. Note
the
  change in absolute scale from left to right.}\label{mdcedc} 
\end{figure*} 

\begin{figure}[h]
\begin{center}
\includegraphics[width=8.6cm]{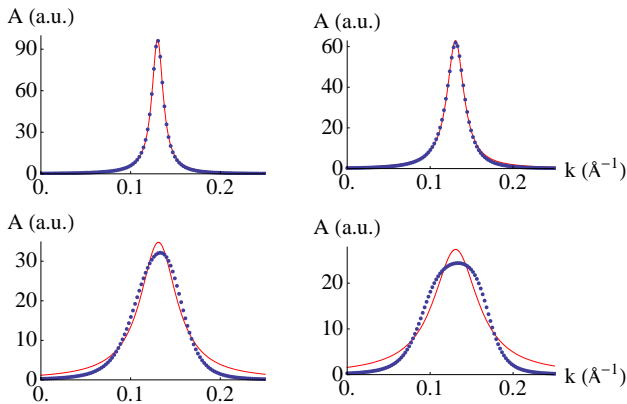}
\caption{MDCs at fixed $\omega = 0.8$ $ eV$ for $n_{imp} = 6 \cdot
  10^{12}$ cm$^{-2}$ and decreasing chemical potential  $\mu = 0.30$ eV (top left),
  $0.20$ eV (top right), $0.09$ eV (bottom left), $0.05$ eV (bottom right). The best Lorentzian fit
to the data is
  shown in red. Note how the lineshape becomes manifestly
  non-Lorentzian as the doping is reduced.}\label{nonlorentz} 
\end{center}
\end{figure}

Finally, we discuss the spectral function of the system, given by
Eq.~(\ref{spec1}). The spectral function is a physical magnitude that is
directly measured in ARPES experiments
\cite{BOM07,ZGF07,BOS07,SSH09}. A customary way of representing the spectral function is through two
sets of plots: its constant $\omega$ sections as a function of $k$, known as momentum distribution
curves (MDCs) and its constant $k$ sections as a function of $\omega$, known as energy distribution
curves (EDCs). Here we show both of them in Fig.~\ref{mdcedc}, for $n_{imp}=3\cdot 10^{12}$
cm$^{-2}$ and $n=2 \cdot 10^{11}$ and $6 \cdot10^{12}$ cm$^{-2}$. We
observe the typical broadening of the quasiparticle Lorentzian peaks
due to disorder, whose width increases with increasing impurity
density. However, a closer look reveals an unexpected
feature. Fig. \ref{nonlorentz} shows the evolution of the MDCs with
decreasing doping, and the best Lorentzian fit to each curve is also
shown. It is appreciated that as doping is decreased the lineshapes
become strongly non-Lorentzian. This is due to the momentum dependence
of the self-energy: the MDCs are obtained from Eq.~(\ref{spec2}) by fixing
the frequency. If the function $\Sigma(k)$ changes significantly from
its on shell value $\Sigma(k_0)$ (with $k_0$ defined by $\omega_0 =
\epsilon(k_0)$) within a scale $\eta$, then regarding the MDCs, the
self-energy looks like a constant if $\Sigma(k_0,\omega_0)<< \eta$,
and therefore the MDC looks like a Lorentzian. Analyzing our numerical
data, it can be seen that this condition is fulfilled only for small
impurity concentrations, and this is the reason for the non-Lorentzian
peaks shown in fig. \ref{nonlorentz}. 

\subsection{The optical conductivity}\label{opticalsec}

In this section we investigate the optical conductivity of graphene,
which is affected strongly by long-range charged impurities.
The optical conductivity of the ideal (or intrinsic) Dirac fermion model is
predicted to be frequency independent and given by the universal value
\cite{CGP09} 
\begin{equation}\label{sigma1}
\sigma_0 \equiv \frac{\pi}{2} \frac{e^2}{h}.
\end{equation}
This frequency independent optical conductivity has been measured in
both suspended samples 
\cite{NBG08} and samples deposited on SiO$_2$ \cite{MSW08}.
This universal result is observed even beyond the energies where the Dirac model
is valid, due to the smallness of
the correction induced by trigonal warping \cite{SPG08}. The absence
of sizable electron-electron interaction corrections in these experiments
has also been recently discussed
\cite{HJV08,M08,SS09}. Moreover, this ideal picture may be modified
by a thermal broadening or a level broadening of
the single particle states due to disorder \cite{PGC06,RMF07}. In the
presence of the broadenings the universal value Eq.~(\ref{sigma1}) 
is obtained only in the asymptotically high energy limit, where all
energy scale of broadenings are negligible. 

Optical conductivity experiments have also been performed in the gated
system \cite{LHJ08,HCG10}, where the optical
conductivity with different electron densities has been studied.
Theoretically, the optical conductivity at finite carrier density (or finite
chemical potential $\mu$) has
been addressed by many authors
\cite{ASZ02,GSC06,PLS07,FP07a,SPC08,GVV09,PRC10,P10a}. Without any disorder
and at $T=0$
the optical conductivity at finite density arises from two contributions:
a Drude peak at zero energy from intraband transitions and a
constant contribution from interband transitions, Eq.~(\ref{sigma1}), starting
at the threshold energy $\omega = 2\mu$.  
However, in the real experiment \cite{LHJ08}, a number of
strong deviations from these ideal predictions have been observed.p10a
One of them is a constant background conductivity in the region
between the Drude peak and 
the threshold at $\omega = 2\mu$, where the interband optical
transition is forbidden. 
An anomalously large energy broadening of the threshold has also been
observed, which cannot be explained in terms of thermal broadening. 
This anomalous optical conductivity in the forbidden region has been
attributed to various physical mechanisms. Short
range impurities produce a broadening of both the Drude peak and the
threshold \cite{ASZ02}, and more recently Coulomb impurities have been
considered within Born approximation to produce a similar but stronger effect
\cite{SPC08}. Electron-electron interactions are also considered 
to explain the background optical conductivity
\cite{GVV09,PRC10}.  In addition, the measured optical conductivity of
very low mobility samples in CVD graphene \cite{HCG10} shows
a reduction of the free carrier Drude 
weight induced by the intraband transition and consequently a 
substantial weight increase due to interband
transition, which can not be explained within Born approximation.

While the effects of short-range scatterers on the optical
conductivity have been computed in the SCBA \cite{ASZ02}, Coulomb
impurities have only been considered within the first order Born approximation
\cite{SPC08}. In this section we calculate the optical conductivity of
graphene within the self-consistent Born approximation and provide the
experimental relevance of our results.
We use the Kubo formula with the SCBA self-energies calculated in the
previous section. 
The optical conductivity can be computed from the current-current
correlation function as
\begin{equation}\label{optcond}
\sigma(\omega) = \frac{g_{s} g_{v}}{2\pi \omega} \int_{-\omega +
  \mu}^{\mu} dE \int kdk \sum_{ss'} A_{s}(k,E)A_{s'}(k,E+\omega) 
\end{equation}
with $A_{s}(k,\omega)$, the spectral function defined in Eq.~(\ref{spec2}).  

\begin{figure}[h]
\begin{center}
\includegraphics[width=8cm]{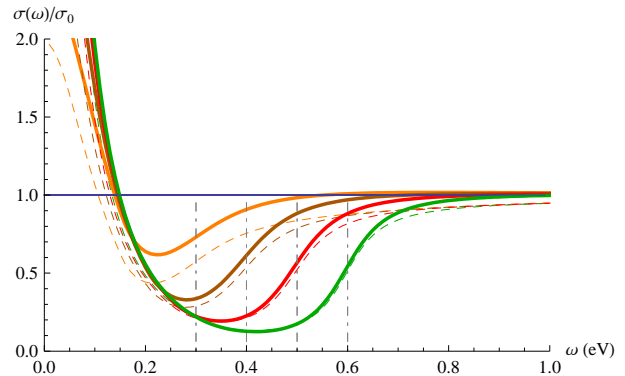}
\caption{Real part of the optical conductivity 
  for a impurity density $n_{imp} = 6\cdot 10^{12}$ cm$^{-2}$ and
  several chemical potentials $\mu = 0.15$, $0.20$, $0.25$, $0.30$ eV.
The solid (dashed) lines indicate the optical conductivities calculated 
  with the SCBA self-energy (with the first order self-energy at $k=0$).
  The corresponding values for $\gamma$ for each curve
  are 0.63, 0.35, 0.22 and 0.16.}\label{optical} 
\end{center}
\end{figure}

Fig.~\ref{optical} shows the optical conductivity as a function of
energy for different electron densities
with the full self-energies
obtained from the self-consistent Born approximation. For comparison,
the optical conductivity with the first order self-energy given by
Eq.~(\ref{firstk0}) is also shown.  
As expected, the result approaches to the
value $\sigma_0$ for $\omega \gg 2\mu$. Most importantly,
the SCBA result gives significantly more background conductivity for
the density $n<n_{imp}$. We note that the four curves
correspond to values of $\gamma$ = 0.63, 0.35, 0.22 and 0.16, for
given chemical potentials of $\mu =$0.15, 0.20, 0.25, and 0.30 eV, respectively.
As discussed in the previous section, 
 SCBA effects are more significant for the
low electron densities. The
self-energy relevant for this computation is the on-shell self-energy,
which is clearly different from the $k$=0 self-energy.
In particular, it gives a non-zero broadening at the Dirac
point, which is partially responsible for the background at $\omega$
close to $\mu$.  

While these results improve on those computed with the first order self-energy
for $n_{imp} > n$, ,
they also indicate that Coulomb impurities are not the only contribution in the
experiment in \cite{LHJ08}. Indeed, any scattering mechanism which is affected
by screening will tend to produce less scattering as the doping is increased,
and what is needed to explain the background in the forbbiden region is a
mechanism which becomes more efficient at high dopings. Our results are
consistent, however, with the latest experiment reported in \cite{HCG10}, which
seems to be dominated mainly by disorder. 

\section{Discussion and conclusions}\label{discussion}

We now discuss several aspects of the results shown in the previous
sections.  We have shown that the $k=0$ self-energy widely employed for the
electron self-energy is only valid near 
the Dirac point and  very different from the
on-shell self-energy. Thus it is necessary to consider the on-shell
self energy to describe the
electronic properties of disordered graphene. 
The on-shell self-energy with bare single particle energy
is not much different from
the true self-energy and represents a reasonable approximation. 
We also discuss the Thomas-Fermi (TF) screening approximation and its
relation with the RPA  approximation. Since the TF
screening function is equivalent to the RPA result for $k<2k_F$, it
seems that the self-energies calculated within both approximations should
coincide at low energies. However, as shown in Fig.~\ref{compsigma}
the TF self-energy differs from its RPA counterpart at all energies.
The difference becomes greater at higher values of $\gamma$ as defined in Eq.~(\ref{gamma}),
especially for high
impurity concentrations. This is because the SCBA  is
non-local in momentum space and the difference between TF and RPA at high
momenta is enough to alter the low-energy region of the
self-energy. It is therefore more reliable to use the RPA screening
function even for $\omega< 2v_F k_F$. 

We now compare our SCBA results with those obtained in the same approximation for short range
disorder
\cite{HHD08,PGC06}. We find that even though
the densities of states for both cases are
similar in energy dependence at low impurity densities the dependence of
both $n$ and $n_{imp}$ is very different.  
We also find that the single particle relaxation time behaves very
differently depending on whether $\gamma >1$ or $\gamma < 1$.
The parameter $\gamma$ 
determines to what extent the SCBA differs from the first
order Born approximation. It is a good consistency test that for small
values of $\gamma$, our numerical integration results match perfectly
with the first order Born approximation Eq.~(\ref{firstborn}). This allows
us to see clearly the deviation from linearity as the disorder density
increases.
This deviation is characteristic of Coulomb impurities which gives
rise to the ratio $\tau_{tr}/\tau_{e}$  to deviate from its
constant value at low impurity densities. 

The momentum distribution curves (MDC)  calculated 
within the self-consistence Born
approximation have also revealed an
unexpected feature.
They become non-Lorentzian with decreasing 
electron density, which suggests that the evolution of MDCs with density could
be
used to probe the relevant scattering mechanisms in graphene. If the
MDC become non-Lorentzian by lowering the electron density, this is a signature
of a momentum dependent self-energy, which could be 
produced by Coulomb impurities (other momentum-dependent potentials
such as ripples could perhaps produce a similar effect, but the method
is still useful to distinguish short range scatterers from long range
scatterers).

Finally, the discussed spectral features have a direct influence on the optical
conductivity. We have shown that long range Coulomb impurities beyond the Born
approximation induce a characteristic broadening in the shape of the optical
conductivity at finite doping compatible with recent experiments.   

In summary, this work has addressed the effects of long ranged charged
impurities on the electronic properties of graphene, showing how
self-consistent scheme modifies them as the impurity density $n_{imp}$
is increased. The SCBA in the presence of Coulomb disorder produces
characteristic  
features in both the spectral properties and the optical conductivity
which may be relevant in the explanation of current experiments.
 
\section{Acknowledgments}
We would like to thank A.G. Grushin and E.V. Castro for very useful
discussions. F.J. was supported by the I3P predoctoral program from
CSIC and from  MICINN (Spain) through grant FIS2008-00124.
EHH acknowledges a financial support from US-ONR-MURI.

\bibliography{zotero}

\end{document}